# Big data need physical ideas and methods

J. P. Huang

Department of Physics, Fudan University, Shanghai 200433, China

*Email: jphuang@fudan.edu.cn*

**If a person looks at WHITE paper through BLUE glasses, the paper will become BLUE in the eye of the person. Likewise, in the current study of big data which play the same role as the white paper being looked at, various statistical methods just serve as the blue glasses. That is, results obtained from big data often depend on the statistical methods in use, which may often defy reality. Here I suggest using physical ideas and methods to overcome this problem to the greatest extent. This suggestion is helpful to development and application of big data.**

With the popularization of computer and Internet, big data come to appear in every aspect of our human lives: they involve many different disciplines and fields [1-19] like physics/astronomy, biology, medicine, industry, economics/finance, politics, education, and so on.

**A. What are big data?**

In the eye of an experimental physicist, the sample of ferrofluids [20] fabricated can be characterized by structure, function, and mechanism bridging the structure and function. Similarly, big data can also be characterized by these three factors, namely, structure, function, and mechanism.

The structure of big data has the feature – "massiveness". What is "massiveness"? For example, before the advent of computer and Internet, my father bought a book in a bookstore; the sales assistant could slowly record the data by writing: one book sold with the price of 0.5 Chinese Yuan on Feb. 25, 1977. But, in the current Internet era, if I buy a book online, the website will quickly record much more data: except for "one book sold with the price of 25 Chinese Yuan on Sep. 14, 2014", more data can be stored immediately, say, the trading time up to the resolution of second, the identity details of me (including gender, birth date, education level, etc.), and so on. Clearly, the data are much more multidimensional and abundant than in the old times, implying a large variety and big volume in view of a huge number of online buyers. So, I may say that the above-mentioned "massiveness" is mainly due to the large variety and big volume of big data. In other words, the "massiveness" is not absolute, but relative.

The function of big data is featured by "valuability". For example, the above-mentioned data about buying book online allow one to analyze and reveal more valuable results (reflecting the complexity of big data), say, predicting human behavior patterns [1, 2]. Thus, "valuability", which is closely related to the complexity of big data, means that the results obtained from big data should be able to both (either) explain the past/knowns and (or) predict the future/unknowns. The requirement of "valuability" is particularly important for big data; or else, it is not necessary for us to call them "big data". For instance, although white noise data could have the feature of "massiveness", the data are not big data due to their uselessness. On the other hand, to make big data valuable indeed, the velocity or veracity for collecting these data should be guaranteed as well.

For big data, revealing the mechanism bridging the structure and function is the most important task of scientists. This is because without a clear understanding of the mechanism, the function revealed in big data may be misleading, unreliable, or even false. So far, many statistical tools have been established and developed to analyze big data [11, 14, 21, 22].

The main theme of the present article is to present how to use the genuine paradigm of scientific research to reveal the mechanism bridging the structure and function of big data. Hopefully, it is helpful to development and application of the big data science.

### B. What is the genuine paradigm of scientific research?

It relies closely on physical ideas and methods. Why? In general, science can be divided into two branches: natural sciences and social sciences. Physics, which focuses on material structures/properties/mechanisms with the level of atoms or smaller, is the fundamental of all the other natural sciences like chemistry and biology. Here, in brief, chemistry is chiefly concerned with the reactions between molecules, and biology involves mainly the activities of cells and organisms. Moreover, currently the key development of social sciences also refers to natural sciences, focusing on quantitative researches. For example, it is often mentioned that within social sciences, economics is the most close to natural sciences since economic data are usually analyzed by mathematical tools, just as natural scientists do [12]. In addition, many economic behaviors accord with theories in physics [4-10, 23-26].

### a. What are physical ideas?

Table 1: Comparison between two physical ideas (PIs)

| PIs | Key pioneers | Role |
|---|---|---|
| **PI 1: Reasons extracted should be coarse-grained** | G. Galilei (Feb. 15, 1564 - Jan. 8, 1642) | Working for reasons |
| **PI 2: Results obtained should be universal** | I. Newton (Dec. 25, 1642 - March 20, 1726) | Working for results |

*(1) Reasons extracted should be coarse-grained*

Let me take the freely falling object as an example. The number of factors determining falling height could be up to *N*: time, air resistance, atmospheric pressure, humidity, dark matter [27], dark energy [28], etc. However, G. Galilei [29] neglected all the *N-1* factors, and considered only the relation between falling height (*h*) and time (*t*), yielding $h = (1/2)gt^2$. Here *g* is acceleration (a constant). As a result, he established the law of free fall, which helped I. Newton [30] to successfully establish classical mechanics in physics. According to this law, the first idea of physics comes to appear: one should extract crucial factors (reasons), or equivalently *reasons extracted should be coarse-grained*.

*(2) Results obtained should be universal*

After Galilei's *h=(1/2)gt$^2$* [29], I. Newton [30] established his second law, *F=ma*, where *F* is force, *m* is mass, and *a* is acceleration. It helps to explain not only the freely falling object on the earth (by setting *a=g* and seeing *F* as gravity), but also the planetary motion in the sky (that had been empirically summarized in the laws of planetary motion by J. Kepler [31]). Besides, Newton's second law can even be used to predict new phenomena. Say, on Aug. 31, 1846, U. Le Verrier (March 11, 1811 - Sep. 23, 1877) [32] first predicted the existence and position of Neptune by using Newton's second law and Newton's law of gravity; Neptune was subsequently observed on Sep. 23, 1846 by J. G. Galle (June 9, 1812 - July 10, 1910) and H. L. d'Arrest (Aug. 13, 1822 - June 14, 1875) [32]. The success of Newton's second law indicates that the second idea of physics is that *"results obtained should be universal"*. Here, the "universal" means that the results should not only help to explain the past or knowns, but also help to predict the future or unknowns.

### b. What are physical methods?

Table 2: Comparison among three physical methods (PMs)

| PMs | Key pioneers | Roles |
|---|---|---|
| **PM 1: Empirical analysis** | Aristotle (384 - 322 B.C.); J. Kepler (Dec. 27, 1571 - Nov. 15, 1630) | Observe/analyze natural phenomena, revealing correlations |
| **PM 2: The combination of empirical analysis and controlled experiments** | G. Galilei (Feb. 15, 1564 - Jan. 8, 1642) | Get crucial reasons for the above phenomena (PI 1), revealing causations |
| **PM 3: The combination of empirical analysis, controlled experiments and theoretical analysis** | I. Newton (Dec. 25, 1642 - March 20, 1726) | Generalize the above causations (PIs 1-2), in order to explain the past or knowns and predict the future or unknowns |

*PM 1: Empirical analysis*

From Aristotle to J. Kepler, physicists first observed the natural world, and then analyzed the observations, yielding many empirical results, say, Kepler's laws of planetary motion [31]. Such analyses are empirical, which are just based on existing data in nature.

Advantages of PM 1 (empirical analysis): reliability and huge data. Here, "reliability" means that according to the data collected from the nature itself, results obtained from the data should be reliable (at least to some extent); "huge data" means that the number of data in nature is huge, which is definitely helpful for understanding the natural world.

Disadvantages of PM 1 (empirical analysis): uncontrollability (correlation) and non-formatting. Since the data are collected from the nature, they are always uncontrollable. Then, what can be obtained from empirical analysis are correlations, but not causations. The latter represent a deeper understanding than the former. On the other hand, it is easy for one to understand "non-formatting" since the format of data existing in nature depends on how people collect them. That is, different people prefer different kinds of formats, thus causing troubles for people to investigate.

*PM 2: The combination of empirical analysis and controlled experiments*

Since empirical analysis helps to reveal correlations rather than causations, G. Galilei [29, 33] started to perform experiments in the laboratory by purposefully tuning one or a few parameters/conditions (but all the other parameters/conditions are fixed), in order to reveal cause-effect relationships (causations). His method is called controlled experiments which are often conducted in the light of empirical analysis, thus yielding PM 2.

Advantages of PM 2: controllability (causation) and formatting. These are just the inverse of the above-mentioned disadvantages of empirical analysis (PM 1). Such experiments are controllable because one can tune one variable and see its effect (causation). Regarding "formatting", it means that the format of data could be conveniently organized during controlled experiments.

Disadvantages of PM 2: deviations and few data. Since such experiments are conducted in the laboratory, the experimental data may be different from their counterparts in nature. The difference is just what I mean "deviations". On the other hand, the experimental data produced in the laboratory cannot be huge, as one can imagine. Thus, I indicate "few data" herein.

*PM 3: The combination of empirical analysis, controlled experiments and theoretical analysis*

Due to the above-mentioned disadvantages of PM 1, I. Newton [30] also started from PM 2; for instance, when he explained Kepler's laws of planetary motion (outcome of empirical analysis), he also explained Galilei's law of free fall (outcome of controlled experiments). The combination of both empirical analysis and controlled experiments (PM 2) reserves their advantages, and removes their disadvantages. More importantly, Newton [30] also realized that the combination of both empirical analysis and controlled experiments (PM 2) can only produce results for specific areas: empirical analysis corresponds to the specific systems producing empirical data (e.g., Kepler's laws of planetary motion are only valid for the planets [31]); controlled experiments are related to specific laboratory samples/devices producing experimental data (say, Galilei's law of free fall [29, 33] specifically holds for the freely falling objects in the laboratory). As a result, Newton [30] utilized theoretical analysis (based on mathematics like calculus) to generalize the results (obtained from PM 2) from specific areas to broad areas. For example, his second law (*F=ma*) can help to not only explain the motion of either planets (described by Kepler's laws of planetary motion [31]) or freely falling objects (described by Galilei's law of free fall [29, 33]), but also predict the motions of many other objects including a single molecule. Owing to the unprecedented success of this generalization (which has been convincingly proved by the fact that physics has helped to improve our lives significantly. For example, satellites, launched according to Newton's second law and Newton's law of gravity, enable us to enjoy satellite TV programs at home), the method of combining empirical analysis, controlled experiments and theoretical analysis (PM 3) has become the fundamental method for developing modern physics. Certainly, it is already enough for achieving some excellent results by using only one or two approaches within empirical analysis, controlled experiments and theoretical analysis. This fact depends on specific topics. For example, in the field of modern condensed matter physics like iron-based superconductors [34], empirical analysis can hardly be adopted. However, in the area of modern astrophysics [35], conducting controlled experiments is almost impossible. It is not necessary for me to go into more details herein. In principle, the aforementioned PM 3 is an ideal, complete method for scientific researches in physics. This conclusion should be sound since physics (involving PM 3) has been developed from the Aristotle time to nowadays and has helped to improve human lives significantly.

So far, I have answered the question "what are physical ideas and methods". The answer has been briefly summarized in Tables 1-2.

## C. What is the big data science?

It means using the above-mentioned physical ideas (PIs 1-2 in Table 1) and methods (PM 3 in Table 2) to investigate big data; see also Table 3.

Table 3: PMs (including PIs) for the big data science

| *PMs* | *Roles* |
|---|---|
| **PM 1: Empirical analysis** | Analyze big data, revealing correlations |
| **PM 2: The combination of empirical analysis and controlled experiments** | Get crucial reasons (PI 1), revealing causations |
| **PM 3: The combination of empirical analysis, controlled experiments and theoretical analysis** | Generalize the above causations (PIs 1-2), in order to explain the past/knowns and predict the future/unknowns |

Clearly, the big data science is equivalent to big data plus the genuine paradigm of scientific research that is composed of physical ideas (PIs 1-2 in Table 1) and methods (PM 3 in Table 2 or 3). From Table 3, it can be readily seen that getting conclusions by purely analyzing big data belongs to empirical analysis (PM 1), which is only at the early stage of PM 3. However, currently PM 1 dominates the literature of big data researches.

### a. Are the results obtained from PM 1 for big data reliable?

To answer this question, I'd like to start from the field of traditional physics: Are the results obtained from PM 1 for natural phenomena reliable?

*(1) An example of failure in physics: "the sun goes around the earth"*

According to empirical observations of natural phenomena like the distribution of stars, Aristotle concluded that "the sun goes around the earth" [36]. His analysis belongs to PM 1; according to PM 3, there still lack controlled experiments and theoretical analysis. Then, Galilei conducted controlled experiments – freely falling object [29, 33] (PM 2), and Newton further gave theoretical analysis – Newton's second law and gravity law [30] (PM 3). As a result of Newton and his followers, the conclusion should inversely be "the earth goes around the sun", which also echoes with the view of N. Copernicus (Feb. 19, 1473 – May 24, 1543) [37]. That is, Aristotle's conclusion based on PM 1 failed to pass the test of PM 3.

*(2) An example of success in physics: "there is no vacuum"*

After scrutinizing the surrounding natural circumstance, Aristotle affirmed that "*there is no vacuum*" [36]. His analysis is just empirical observations (PM 1); in view of PM 3, both controlled experiments and theoretical analysis are still in need. Accordingly, experimentalists performed controlled experiments: high-energy photons excite lots of particles within a space where all kinds of matters (that people can imagine, say, air) have been driven off [38] (PM 2). Also, theorists established the theories: the Dirac equation and associates [39, 40] (PM 3). As a result, it is concluded that there is no vacuum indeed. Namely, Aristotle's conclusion according to PM 1 passed the test of PM 3. (Incidentally, it is worth mentioning that owing to historical limitations, Aristotle's "no vacuum" [36] has other specific connotations that are out of the scope of this article.)

In fact, regarding the above question "Are the results obtained from PM 1 for big data reliable", its answer is quite similar to that in the field of physics. In the following, I shall also give two examples: one is a failure,

**14 / 24**

the other is a success.

*(3) An example of failure in big data: "the risk-return relationship is positive"*

The empirical analysis (PM 1) helped scientists to consider investments as high risk high return and vice versa [41, 42]. That is, the risk-return relationship is positive (risk-return tradeoff) [41, 42], which is the majority view in the literature on risk management. However, PM 3 implies that the analysis method is still short of controlled experiments and theoretical analysis. Therefore, experimentalists conducted controlled experiments; but they revealed an inverse result: when markets have the features of both efficiency and closeness, the risk-return relationship is negative (namely, high risk low return and vice versa) [23,43]. Further the theoretical analysis confirmed the experimental findings [23,43]. Such findings echo with Bowman's paradox [44, 45], which is the minority view in the literature on risk management. In other words, the aforementioned majority view (namely, "the risk-return relationship is positive") obtained from PM 1 does not completely stand the test of PM 3.

*(4) An example of success in big data: "markets have an invisible hand"*

A. Smith (June 16, 1723 – July 17, 1790) empirically considered various

kinds of market phenomena (PM 1), and asserted that *markets have an invisible hand*, which tends to drive markets to equilibrium (namely, the balance between supply and demand) [46]. Clearly, owing to PM 3, his analysis still needs controlled experiments and theoretical analysis. So, experimentalists conducted a relevant kind of controlled experiments for allocating resources with unbiased or biased distributions [23,24] (PM 2), and theorists further gave theoretical analysis based on agent-based simulations [23,24] (PM 3). Finally, the controlled experiments and theoretical analysis help to provide strong evidence for the existence of "invisible hand". Namely, Smith's assertion originating from PM 1 stands the test of PM 3.

### D. Discussion and conclusion

Studying big data should obey the genuine paradigm of scientific research, which is composed of physical ideas (PI 1-2 in Table 1) and methods (PM 3 in Table 2 or 3). It is not enough to analyze big data by using PM 1 only; it is the most appropriate way to use PM 3, in order to discover useful rules/laws behind big data. Certainly, when a person cannot use his/her two legs to walk smoothly, only one leg is also okay for "walking" even though his/her walk is not so smooth. That is, if PM 3 cannot be used for handling big data due to the limitation of some real situations, either PM 1 or PM 2 can also be performed initially as a compromise. Doing so can help people to develop science as much as possible. Or else, the development of science will stop.

On the other hand, someone may ask me a question: "we know the importance of PM 3, but can you tell me how to realize PM 3 in practical big data researches?" Frankly, I cannot give a general answer to the question because the answer must rely on particular cases. But, in the light of both the success of PM 3 in physics and the two examples introduced above [23-26,41-46], I am confident that the outcomes of PM 3 in the big data science are worth waiting for. That is, further researches are needed, following the above-listed two examples.

So far, for big data, one might have kept it in mind that PM 3 could yield more general results than PM 2, and that PM 2 can produce more insightful results than PM 1. For PM 1, here I would like to recommend the reader to read an excellent review [3], which presents not only the statistical tools which yield unreliable results for big data analysis (say, noise accumulation [47] and spurious correlation [48]), but also more comprehensive statistical methods [21, 22] for improving the reliability of big data analysis (say, penalized quasi-likelihood [49] and independence screening [48]). Clearly, using such statistical tools/methods still belongs to PM 1 as discussed in this article.

To sum up, if a person observes white paper through blue glasses, the paper becomes blue in the eye of the person; when one uses statistical tools (PM 1) to treat big data, the big data serve as the "white paper" being observed, and the statistical tools (PM 1) adopted may often serve as the "blue glasses". To make current big data analysis useful to the greatest extent, I suggest to utilize PM 3, which has been demonstrated to be powerful in both physics (the most fundamental discipline in natural and social sciences) and some examples in big data [23-26,41-46].

## Acknowledgements

I acknowledge the financial support by the National Natural Science Foundation of China under Grant No. 11222544, by the Fok Ying Tung Education Foundation under Grant No. 131008, by the Program for New Century Excellent Talents in University (NCET-12-0121), and by Shanghai Key Laboratory of Financial Information Technology (Shanghai University of Finance and Economics).